\documentclass[prb,twocolumn,preprintnumbers,superscriptaddress]{revtex4-1}
\usepackage{graphicx}% Include figure fiiles
\usepackage{dcolumn}% Align table columns on decimal point
\usepackage{bm}% bold math
\usepackage{epsfig}

\usepackage{longtable}

\begin{document}

\preprint{Submitted to {\it Review of Scientific Instruments}; LA-UR-12-20616}

\title{Gas gun shock experiments with single-pulse x-ray phase contrast imaging and diffraction at the Advanced Photon Source
}

\author{S. N. Luo}
\thanks{sluo@lanl.gov}
\affiliation{
Los Alamos National Laboratory, Los Alamos, New Mexico 87545, USA
}

\author{B. J. Jensen}
\affiliation{
Los Alamos National Laboratory, Los Alamos, New Mexico 87545, USA
}

\author{D. E. Hooks}
\affiliation{
Los Alamos National Laboratory, Los Alamos, New Mexico 87545, USA
}

\author{K. Fezzaa}
\affiliation{
Advanced Photon Source, Argonne National Laboratory, Argonne, Illinois 60439, USA}

\author{K. J. Ramos}
\affiliation{
Los Alamos National Laboratory, Los Alamos, New Mexico 87545, USA
}

\author{J. D. Yeager}
\affiliation{
Los Alamos National Laboratory, Los Alamos, New Mexico 87545, USA
}

\author{K. Kwiatkowski}
\affiliation{
Los Alamos National Laboratory, Los Alamos, New Mexico 87545, USA
}

\author{T. Shimada}
\affiliation{
Los Alamos National Laboratory, Los Alamos, New Mexico 87545, USA
}

%\author{D. A. Fredenburg}
%\affiliation{
%Los Alamos National Laboratory, Los Alamos, New Mexico 87545, USA
%}

%\author{D. M. Dattelbaum}
%\affiliation{
%Los Alamos National Laboratory, Los Alamos, New Mexico 87545, USA
%}

\date{\today}

\begin{abstract} 
The highly transient nature of shock loading and pronounced microstructure effects on dynamic materials response call for {\it in situ}, temporally and spatially resolved, x-ray-based diagnostics. Third-generation synchrotron x-ray sources are advantageous for x-ray phase contrast imaging (PCI) and diffraction under dynamic loading, due to their high photon energy, high photon fluxes, high coherency, and high pulse repetition rates. The feasibility of bulk-scale gas gun shock experiments with dynamic x-ray PCI and diffraction measurements was investigated at the beamline 32ID-B of the Advanced Photon Source. The x-ray beam characteristics, experimental setup, x-ray diagnostics, and static and dynamic test results are described. We demonstrate ultrafast, multiframe, single-pulse PCI measurements with unprecedented temporal ($<$100 ps) and spatial ($\sim$2 $\mu$m) resolutions for bulk-scale shock experiments, as well as single-pulse dynamic Laue diffraction. The results not only substantiate the potential of synchrotron-based experiments for addressing a variety of shock physics problems, but also allow us to identify the technical challenges related to image detection, x-ray source, and dynamic loading.   
\end{abstract}
\maketitle

\section{Introduction}
Dynamic loading rate and path as well as microstructure play vital roles in dynamic materials response. Designing materials that function at dynamic extremes and predicting their response require broad and in-depth experimental investigations of the time, rate and microstructure dependences.\cite{davison75pr,meyers94b,kanel10ijf,marie} Key to such experiments are {\it in situ}, temporally and spatially resolved measurements, which are inherently challenging at the temporal and spatial scales of interest. Theoretical development and predictive modeling are undermined by the paucity of such measurements.\cite{marie}

Over the past 60 years, optical techniques such as point, line and two-dimensional (2D) velocity or displacement interferometers have been perfected and contributed greatly to shock compression science, but are mostly limited to surface measurements.\cite{barker72jap,trott01sccm,paisley08rsi,2dvisar} Multiframe, dynamic, charged-particle radiography in particular proton radiography (pRad) is an important development for shock physics given the high penetration power of protons.\cite{morris12ropp,king99nima,dimonte11prl}  
Diagnostics utilizing high energy and high flux x-ray photons and coherent photons such as radiography, phase contrast imaging (PCI), diffraction, and diffraction imaging offer the promise of real time, {\it in situ} measurements.\cite{wang05srn,wilkins96nat,fezzaa08prl,klantar05prl,ice11sci,chapman06np} Advances in synchrotron light sources and dynamic loading provide unique opportunities for shock physics. For example, dynamic x-ray diffraction during powder-gun loading was measured with monochromatic x-rays at the Advanced Photon Source (APS).\cite{turneaure09jap} Recently, we performed static high-speed PCI measurements on low-$Z$ materials,\cite{yeager12comp} and briefly reported ultrafast dynamic PCI measurements under gas gun loading at APS.\cite{jensenaip} Synchrotron x-ray radiography experiments achieved a time resolution of 200 ns and a spatial resolution of 200 $\mu$m on low-$Z$ materials under explosive loading.\cite{evdokov07nima}

Dynamic measurements with single x-ray pulses ($<$100 ps) present challenges to x-ray sources, detectors and dynamic loading, and their integration. In this work, we present technical details on dynamic PCI and diffraction under gas gun loading, with single x-ray pulses at APS. We describe the x-ray beam characteristics related to dynamic experiments in Sec.~II, and the experimental setup including timing and synchronization, detectors and scintillators, and some representative static and dynamic PCI and diffraction results in Sec.~III. Sec.~IV addresses the scientific opportunities and technical challenges, followed by summary in Sec.~V.       

  %%Fig1
\begin{figure}[t]
\includegraphics[scale=0.55,clip]{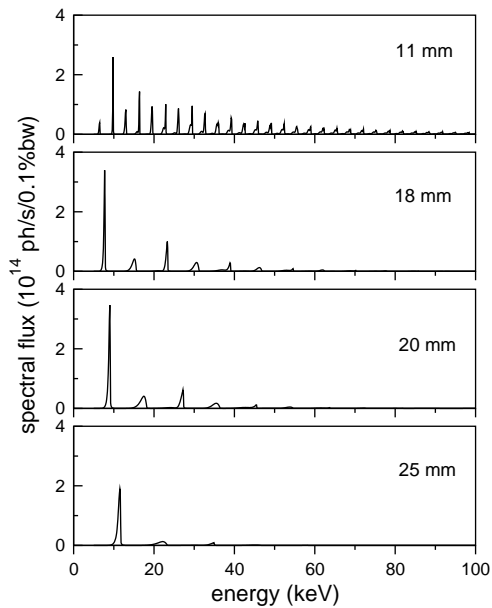}
\caption{The simulated ``white beam'' spectral photon flux through a 1 mm$\times$1 mm pinhole located at 35 m away from the source, for APS Undulator A with different undulator gaps (11 mm, 18 mm, 20 mm and 25 mm). The electron energy is 7 GeV and the current is 100 mA. A monochromatic x-ray beam is also available if Si(111) crystals are used. }
\end{figure}

\section{The x-ray characteristics at the APS beamline 32ID-B}

Gas gun shock wave experiments are highly transient with durations of 100 ns--1 $\mu$s, and are not as exactly reproducible as the laser pump-probe experiments. The samples under investigation normally have dimensions of 1--10 mm. Such experiments require the probing x-ray beam to have sufficiently high photon energy to achieve adequate penetration depth, high photon flux to accommodate the short event duration and reach single-pulse temporal resolution, and appropriate time structure to synchronize the pulsed loading, x-ray beam, and detectors. 
%We thus first discuss the x-ray beam at the APS beamline 32ID-B, where our synchrotron shock experiments have been conducted.

The x-ray beam at 32ID-B employs APS Undulator A with a period of 3.3 cm and length of 2.4 m, with the capability of providing ``white'' or monochromatic x-rays. In order to maximize the number of photons in a single x-ray pulse, the white beam option was selected. The specimen is located approximately 40 m away from the undulator light source. The beam cross-section is of the elongated 2D Gaussian shape; the central region has the highest intensity and may cause damage (heat load) to the objects downstream in the beam path. The FWHM (full width at the half maximum) beam size at 10 keV is 0.6 mm (vertical)$\times$1.3 mm (horizontal).\cite{shen07nima} The actual beam spot size on the sample is controlled with adjustable slits in both directions.

The undulator gap ($g$) is changeable and the photon flux spectral density ($f$) varies accordingly. Fig.~1 shows the representative photon flux spectral densities at different undulator gaps, simulated with XOP and APS Undulator A beam parameters.\cite{xop,apsbeam} The fundamental and high harmonics span over the photon energy ($E$) range of 5--100 keV. With increasing undulator gap, the amplitudes of the high harmonics decrease rapidly. The dominant peak is the fundamental for $g\ge18$ mm.

%%Fig2
\begin{figure}[t]
\includegraphics[scale=0.5,clip]{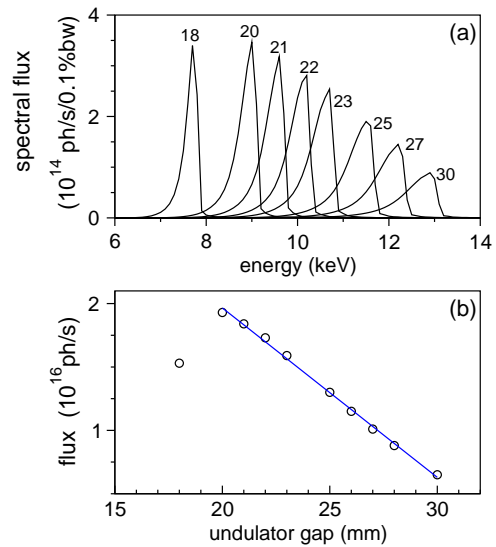}
\caption{(a) The fundamentals of the spectral photon fluxes through a 1 mm$\times$1 mm pinhole located at 35 m away from the source for different undulator gaps (18--30 mm), and (b) the corresponding energy-integrated fluxes. The numbers in (a) denote undulator gap in mm. The line in (b) denotes a linear fit, $y=4.64-0.13x$.}
\end{figure}

Since the fundamental dominates its high harmonics in the spectral photon flux for $g$$\ge$18 mm, and contributes the most to the features in the intended measurements (PCI and diffraction), we plot only the fundamentals in Fig.~2(a) for different undulator gaps. 
For $g$=20--30 mm, the amplitude of the fundamental decreases and the peak broadens with increasing $g$. The peak shifts to higher photon energy from about 9 keV to 13 keV. 
Integrating the photon flux spectral density $f(E)$ over a fundamental yields the photon flux, $F=\int f(E){\rm d}E/E$,
as contributed by the fundamental [Fig.~2(b)]. Here the photon flux refers to that through a 1 mm$\times$1 mm pinhole located at 35 m away from the source. For example, $F$ due to the fundamental is about 1.5$\times$10$^{16}$ ph\,s$^{-1}$ for $g$=18 mm, compared to 2.4$\times$10$^{16}$ ph\,s$^{-1}$ for the full energy range. The photon flux due to the fundamental decreases with increasing undulator gap for $g$$\ge$20 mm [Fig.~2(b)].
% $F$ increases by a factor of 3 if $g$ is decreased from 30 mm to 20 mm, and by a factor of 1.5 from 25 mm to 20 mm [Fig.~2(b)]. 
We can thus change the undulator gap for different photon energies or photon fluxes as needed. The bandwidths of the fundamentals are $\sim$5\% for $g$=20--30 mm, appropriate for PCI (and Laue diffraction). 
  
\begin{table*}[ht]
%\begin{center}
\caption{APS electron storage ring operation modes (total current of 102 mA; 3.68 $\mu$s per one revolution)$^{\rm a}$.}
\begin{tabular}{lllll}
\hline\hline
Mode         & \# of bunches & Bunch length & Bunch current & Bunch separation  \\\hline
Standard   & 24 singlets    &  33.5 ps & 4.25 mA  & 153.3 ns between singlets\\
324-bunch  & 324 singlets    &  22 ps   & 0.31 mA  & 11.37 ns between singlets\\
1296-bunch & 1296 singlets    &  22 ps   & 0.079 mA & 2.84 ns between singlets\\
Hybrid (1+8$\times$7)   & 1 singlet    &  50 ps   & 16 mA    & 1.594 $\mu$s on each side from septuplets\\
                        & 8 septuplets &  27 ps   & 11 mA    & 51 ns between groups\\

\hline\hline
\end{tabular}
\footnotetext[0]{\hspace{1cm}$^{\rm a}$ http://www.aps.anl.gov/Accelerator\_Systems\_Division/Accelerator\_Operations\_Physics/SRparameters/node5.html}
%\end{center}
\end{table*}

The time structure of x-ray pulses or the corresponding electron bunches depends on the operation modes of the APS electron storage ring (Table~I). The pulse train is circular, and one revolution of electrons takes 3.68 $\mu$s. The number of photons in an x-ray pulse scales linearly with the bunch current. The bunch separation can be varied from 2.84 ns to the $\mu$s time scale (Table~I). For time-resolved, single-pulse measurements, we need to consider the bunch current and bunch separation, as well as synchronizing an x-ray pulse with a dynamic event, and the inherent time constants of scintillators and detectors. For example, the singlets in the hybrid mode can supply the most x-ray photons, but synchronizing an x-ray pulse with a dynamic event during gas gun loading would be extremely difficult since the time scale in our initial shock loading experiments is on the order of 100 ns, considerably shorter than the singlet separation (3.68 $\mu$s). The reported single x-ray pulse duration is 80 ps for APS standard mode,\cite{miceli_pilatus} which defines the temporal resolution of single-pulse measurements. Compromising on the temporal resolution, the measurements can be integrated over multiple x-ray pulses. For instance, a temporal resolution of 113 ns can be achieved with 10--11 pulses in the 324-bunch mode. Due to the detector and loading constraints as discussed below, the 24-bunch mode (or the standard APS operation mode) is the most suitable for single-pulse x-ray measurements during gas gun loading, and has been used in our experiments. However, the versatility of the APS x-ray time structure can be fully exploited in the future as the detection and loading technologies advance. For example, the hybrid mode could be utilized for powder-gun loading since the loading jitter is much less than gas-gun loading.

\section{Gas gun experiments with single-pulse x-ray PCI and diffraction}

\subsection{Experimental setup for gas gun shock experiments at APS}

Fig.~3 shows a schematic and a photo of the actual experimental setup for our gas gun shock experiments with x-ray PCI and diffraction at the APS beamline 32ID-B. The ``white beam'' x-rays from the undulator source pass sequentially a slow shutter, a fast shutter, a 2D slit, the sample and detection systems. The shutters are used to control the x-ray-open time window [or simply the x-ray time window; Fig.~4(a)]; the slow shutter is water cooled so it can take high heat load. The 2D slit controls the PCI field of view ($\sim$1--2 mm) on the sample, as well as the sampling spot size ($\sim$0.5 mm) for diffraction. 

%%Fig3
\begin{figure}[t]
\includegraphics[scale=0.56,clip]{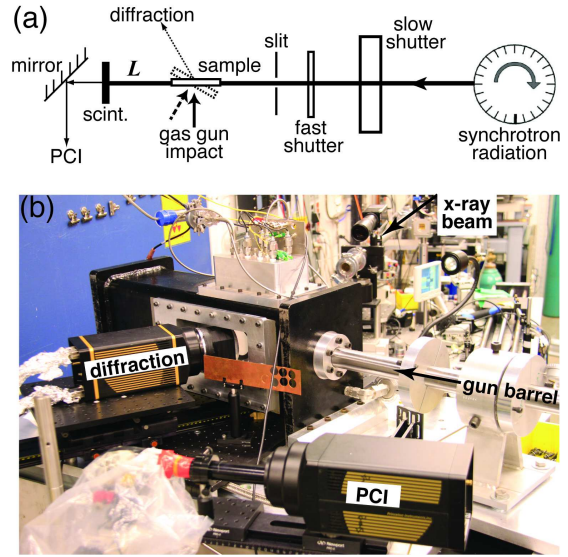}
\caption{(a) Schematic setup for synchrotron shock experiments with x-ray phase contrast imaging (PCI) and diffraction. The sample--scintillator distance $L$ needs to be optimized for best phase contrast in PCI for different materials. (b) A photograph of the experimental setup at the APS beamline 32ID-B (the current gun--x-ray beam configuration is for diffraction in the backscattering mode). Both the sample and gun system can be rotated for different geometry as required by PCI or diffraction: the shock loading direction is perpendicular to the x-ray beam for PCI and transmission-mode diffraction, while it is at an angle (e.g., 10$^{\circ}$--30$^{\circ}$) with the beam for backscattering-mode diffraction.}
\end{figure}

%%Fig4
\begin{figure}[t]
\includegraphics[scale=0.48,clip]{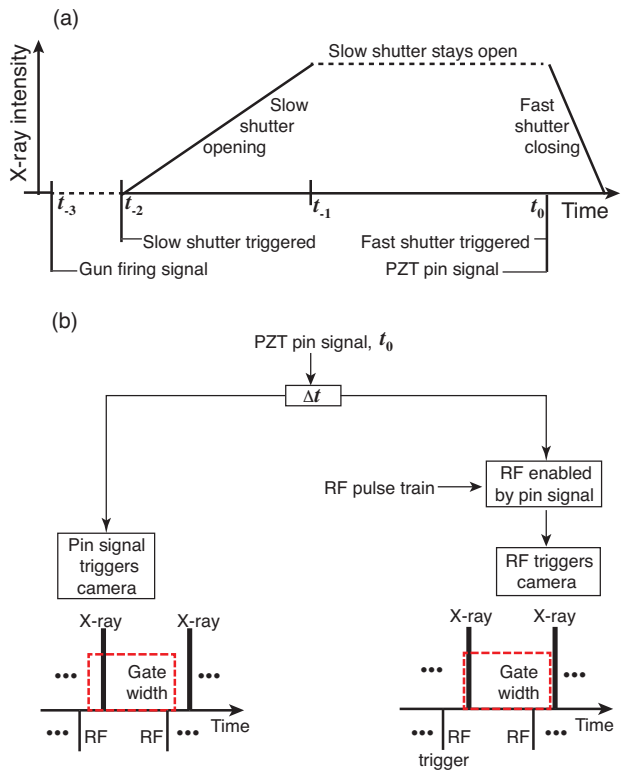}
\caption{(a) Schematic of the time sequence for the x-ray time window. (b) Two schemes for triggering PCI or diffraction cameras to capture a dynamic event occurring at ($t_0+\Delta t$ + intrinsic delays), within the camera gate width (which equals to the x-ray pulse separation).  }
\end{figure}

Materials are subjected to impact loading using a 12.6-mm bore gas gun capable of achieving velocities up to 1 km\,s$^{-1}$. This gas gun is designed specifically for use at a synchrotron source.\cite{jensengun} The gun system consists of a gas breech, a launch tube (or barrel), and a target chamber all mounted on a mobile support structure to allow for insertion and alignment within the x-ray beam. The x-ray beam enters through a side port, is transmitted through a sample, and exits through a second side port where the detector system is located. The side ports are sealed using Lexan windows (approximately 0.25 mm thick) to allow the x-rays to pass through while maintaining vacuum prior to the experiment. During the experiment, the projectile is accelerated down the launch tube and impacts the target, and dynamically compresses the sample. Projectile velocity and surface velocities are measured with multiple probes of standard photonic Doppler velocimetry (PDV). The typical diameter of PCI samples is about 9 mm and the thickness is 1 mm.

The {\it in situ}, time-resolved x-ray diagnostics at APS are ``white beam'' PCI\cite{wilkins96nat,snigiriev95rsi,cloetens96jpd,nugent96prl,kitchen04pmb,yeager12comp} and diffraction, and the 2D PCI/diffraction images are obtained with single x-ray pulses in the APS standard operation mode (24-bunch; Table~I). Diffraction measurements adopt either the backscattering [e.g., Fig.~3(a)] or transmission mode. For PCI and transmission-mode diffraction, the shock loading direction is perpendicular to the x-ray beam, while it is at an angle of 10$^{\circ}$--30$^{\circ}$ with the beam for backscattering-mode diffraction.    

The highest temporal resolution of x-ray PCI or diffraction measurements achievable in dynamic experiments is the width of the probing x-ray pulse, and it is $<$100 ps (80 ps)\cite{miceli_pilatus} for APS standard mode. Within a pulse width of 100 ps, the position change of a shocked sample is on the order of 0.1--1 $\mu$m if we assume a velocity of 1--10 km\,s$^{-1}$, resulting in negligible motion blur. For single-pulse measurements, it is critical to synchronize an x-ray pulse and diagnostics to a dynamic event of interest. In addition, the high flux x-ray beam can exert considerable heat load on an object downstream and may lead to its damage. Therefore, the x-ray beam path between the sample and the x-ray source should remain cleared only for a controllable time window (the x-ray time window). This time window be realized with a normally-open and a normally-closed shutter. It would be convenient to use the radio frequency (RF) pulse train supplied by the synchrotron for triggering and synchronizing the loading, diagnostics and x-ray shutters, but this scheme is unrealistic given the large jitter of the gas gun loading (on the order of 10 ms). Instead, the firing signal from the gun control system and impact PZT (lead zirconate titanate) pin signal was used, along with optional RF signals, for our timing purposes. 

Fig.~4 details the timing scheme for the x-ray time window [Fig.~4(a)] and synchronization of diagnostics to a dynamic event (and the x-ray time window) [Fig.~4(b)]. The gun control system sends the gas valve a firing signal (12 VDC) at time $t_{-3}$ to launch the projectile. This signal also triggers the normally closed, slow shutter to open at time $t_{-2}$ after certain delay (e.g., 100 ms); it takes the slow shutter about 8 ms to completely open at time $t_{-1}$. The accelerated projectile then impacts a PZT piezoelectic pin (Dynasen Inc.) near the muzzle at time $t_0$, and the pin signal triggers the normally open fast shutter to close (the full closure takes about 1 ms). Since the whole loading duration (100 ns--1 $\mu$s) is much shorter than the closing time of the fast shutter (1 ms), the x-ray intensity change during shutter closing is negligible. The x-ray time window is wide (e.g., 60--100 ms) in order to accommodate the jitter in the projectile launch and acceleration.

The PZT pin signal is delayed by $\Delta t$ with a DG535 delay and pulse generator (Stanford Research Systems), and directly or indirectly triggers the PCI or diffraction cameras (and other diagnostics such as PDV) to capture dynamic events [Fig.~4(b)]. For each frame, the camera gate width is equal to the x-ray or RF pulse separation (153 ns for the APS standard mode). There are two ways of triggering the cameras: PZT pin signal trigger or RF pulse trigger. In the former case (pin trigger), an x-ray pulse falls randomly within the camera gate width, and the camera captures a PCI or diffraction image due to this pulse. In the latter case (RF trigger), both the pin signal and RF pulse train are input into a DG535. The RF trigger is enabled when the pin signal arrives at the ``trigger inhibit'' input of DG535, and the first RF pulse after this enabling signal triggers the camera which captures the first x-ray pulse following this RF pulse. Since the phase difference between the RF and x-ray pulses is fixed, the camera can be synchronized to the x-ray pulses to maximize the signal on the camera. This synchronization can be achieved via adjusting the delay between the RF signal and the camera until the maximum signal intensity on the camera is found. In either trigger scheme, the arrival of the x-ray pulse winthin the camera gate width could be off by 1 pulse separation from the trigger. Compared to linear accelerator based light or charged particle sources, APS-type synchrotrons are indeed advantageous in synchronization since the beam is ``continuously'' available and the pulse separation is small. We have mainly utilized the PZT pin trigger scheme, and the RF trigger remains to be fully explored.

%%Fig5
\begin{figure}[t]
\includegraphics[scale=0.43,clip]{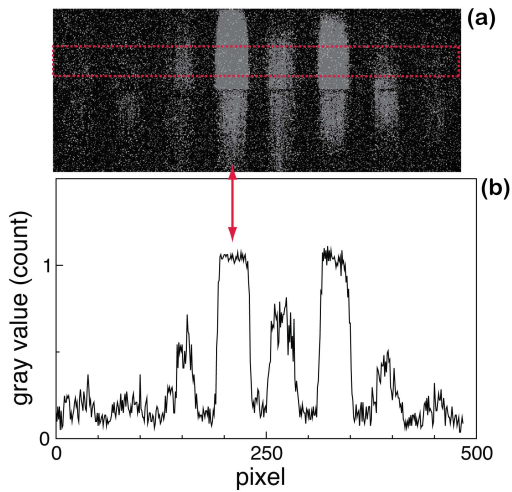}
\caption{(a) ``White beam'' diffraction pattern of a polycrystalline Fe foil as recorded by the Pilatus 100K camera. (b) The intensity profile for the region enclosed by the rectangle in (a). Undulator gap: 40 mm; camera gate width: 40 ns; no scintillator.}
\end{figure}

%%Fig6
\begin{figure}[ht]
\includegraphics[scale=0.42,clip]{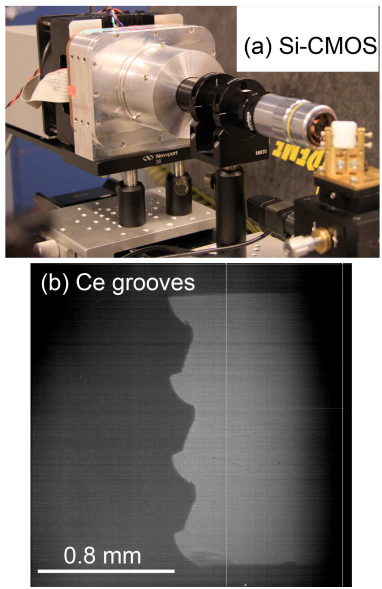}
\caption{(a) The LANL-Rockwell three-frame, hybrid Si-CMOS camera. (b) A raw image (three in total) of a Ce sample with engineered grooves, obtained in a single-pulse static test using the hybrid Si-CMOS camera. Gate width: 140 ns; undulator gap: 26 mm; scintillator: LuAG.
%max 450 counts vs. background 380. LuAG. gain: 3$\times$
}
\end{figure}

%%Fig7
\begin{figure}[ht]
\includegraphics[scale=0.55,clip]{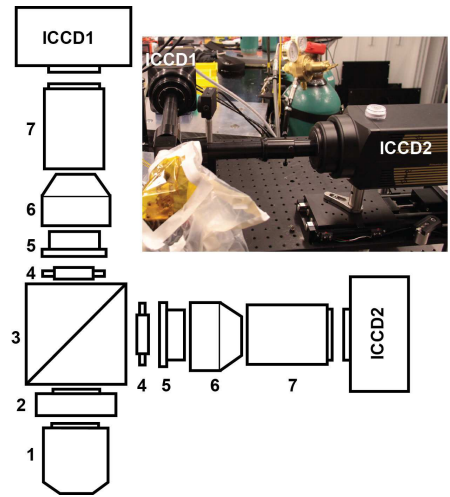}
\caption{Schematic of multiplexing single ICCD cameras. Inset: a photograph of the ICCD multiplexing setup. 1: Mitutoyo infinity-corrected long working distance objective (e.g., 10$\times$, working distance 33.5 mm, depth of focus 3.5 $\mu$m; Edmund Optics NT46-144); 2: Mitutoyo-to-C-mount objective adaptor (Edmund 55-743); 3: C-mount standard cube 50R-50T beam splitter (Edmund 54-823); 4: double male C-mount adaptor (Edmund 03-629); 5: M1 tube lens (Edmund 54-774); 6: M1/M2 tube lens adaptor (Edmund 58-329); 7: 190 mm C-mount extension tube (Edmund 54-631, 54-632 and 54-633). ICCD1 and ICCD2 are PI-MAX ICCD C-mount cameras (Princeton Instruments).}
\end{figure}

%%%%%%%%%%%%%%%%%%%%%%%%%%%%%%
%%%%%%%%%%%%%%%%%%%%%%%%%%%%%%

\subsection{Cameras and scintillators}

Dynamic PCI or diffraction measurements with an x-ray pulse train require that the 2D imaging/diffraction detectors be fast, sensitive, externally triggerable and gateable.  There are indirect- and direct-detection detectors or cameras. The indirect-detection cameras utilize scintillators or phosphors to convert x-ray photons to visible light, which can be optically manipulated and recorded with optical cameras. Optical intensified charge-coupled device (ICCD) cameras can record multiple frames  with light splitting onto multiple cameras, and such framing capability is highly desirable for shock experiments. However, a scintillator's light yield, decay time and resistance to x-ray heat load are limiting factors for the indirect-detection cameras. The multiframe capability is also limited by available x-ray photon fluxes. 

Direct detection multiframe x-ray cameras use semiconductor sensors such as Si and complementary metal oxide semiconductor (CMOS) integrated circuits (or application-specific integrated circuit) to collect x-ray or optical/visible photons. Such detectors include digital and analog pixel array detectors (PADs),\cite{miceli_pilatus,Koerner_pad} and hybrid photosensor-CMOS framing camera.\cite{krisk_camera,hybrid_cmos} Single-photon counting digital PADs are commercially available (e.g., Pilatus).\cite{miceli_pilatus} Photon-integrating analog PADs\cite{Koerner_pad} under development at Cornell University have a 150 $\mu$m$\times$150 $\mu$m pixel size and 16$\times$16 pixels, and are capable of onboard storage of 8 frames with a minimum frame separation of 100 ns. The Los Alamos National Laboratory (LANL) hybrid Si-CMOS camera is similar to the Cornell analog PADs but has larger pixel number and smaller pixel size.\cite{hybrid_cmos} The direct-detection cameras do not use any scintillators, but their damage resistance to high x-ray fluxes remain to be explored. One disadvantage of direct detection is that it is more difficult to manipulate an x-ray beam than an optical beam for magnification, splitting and other purposes. 

The cameras tested in our experiments include the Pilatus 100K camera supplied by the APS detector pool, LANL hybrid Si-CMOS camera, and single-frame PI-MAX ICCD cameras (Princeton Instruments). The scintillators used include LuAG (Lu$_3$Al$_5$O$_{12}$:Ce), LSO (Lu$_2$SiO$_5$:Ce) or LYSO (Lu$_{2-2x}$Y$_{2x}$SiO$_5$:Ce), about 100 $\mu$m thick. These cameras and scintillators are used in static or dynamic PCI or diffraction measurements at the beamline 32ID-B with ``white beam'' single x-ray pulses in the APS standard mode (24 bunches; pulse separation 153 ns). The camera gate width is less than or equal to the pulse separation (153 ns) so the camera is exposed to only a single x-ray pulse. 
%We present below some camera or scintillator details and their tests (Figs.~5--13).     

The Pilatus 100K direct-detection x-ray camera has a 172 $\mu$m$\times$172 $\mu$m pixel size, and a detection area of 83.8 mm$\times$33.5 mm or 487$\times$195 pixels.\cite{miceli_pilatus} Fig.~5 shows a backscattering diffraction pattern of a polycrystalline Fe foil under ambient conditions, obtained with a narrow camera gate width of 40 ns (without a scintillator). High background and saturated diffraction rings are observed. While multiple diffraction rings are evident [Fig.~5(a)], the peak profile (shape and intensity) can not be accurately measured [Fig.~5(b)]. For example, the peak indicated by the arrow is top-hatted, and the count is 1. The low photon-counting rate (up to $\sim$1 MHz)\cite{miceli_pilatus} or long ``deadtime'' leads to the extremely narrow dynamic range for each pixel detector.\cite{Koerner_pad} This single photon counting rate would be too slow for our experiments, where many photons are received by the camera within a short gate width (e.g., 150 ns). Thus, the Pilatus camera was unsuitable for  dynamic measurements that require high photon fluxes.    

The LANL hybrid Si-CMOS camera [Fig.~6(a); fabricated by Rockwell Scientific Company] contains a 720$\times$720 array of 100 $\mu$m thick, fully depleted, Si photosensors and a CMOS readout integrated chip. The latter turns the photosensor signals on and off to measure the intensity of the incident light, and then processes the photosensors' outputs and combines them into a frame. The camera can capture three frames when triggered. The minimum integration time is 130 ns per frame and the interframe separation is 250 ns. The photosensor array and all the control--processing circuitry are contained in a $\sim$20 mm$\times$20 mm hybrid chip. The pixel size is 26 $\mu$m$\times$26 $\mu$m. The quantum efficiency is high, e.g., 84\% at 420 nm and 92\% in the green. A single-pulse, static PCI test was conducted on this camera with an LuAG scintillator and a Ce sample under ambient conditions. One of the three raw frames is shown in Fig.~6(b). While the low intensity can be improved with a reduced undulator gap, the overall image quality as tested is sufficient for high contrast PCI experiments such as the formation of metal jets. 
Improved hybrid Si-CMOS cameras appear to be highly promising for future dynamic experiments.  
 
The Princeton Instruments PI-MAX ICCD optical cameras (e.g., ICCD1 or ICCD2 in Fig.~7 inset) use microchannel plates as light signal intensifiers, and have 1340$\times$1300 pixels with a pixel size of 20 $\mu$m$\times$20 $\mu$m. Each ICCD camera is triggered and gated externally, but can only record one frame when triggered. Fig.~8 shows some examples of static single-pulse x-ray PCI images obtained with a PI-MAX camera (the maximum count is $\sim$3000 at a signal gain of 200$\times$). For multiframe measurements with such ICCD cameras, we constructed a prototype two-frame ICCD camera with two single-frame ICCD cameras and light splitting (Fig.~7; ICCD multiplexing). ICCD1 and ICCD2 are triggered sequentially, separated by one or more x-ray pulse separations. The single- or two-frame ICCD cameras have been used for both static and dynamic PCI and diffraction measurements as shown in Figs.~8--13; the dynamic results will be addressed in more details in the next subsection. 

For diffraction, a 2:1 fiber optic taper (Incom, Inc.) is mounted to the front end of the ICCD camera to quadruple the coverage area of the diffraction signals. The diameter is 75 mm at the large end of the taper and 40 mm at the small end, and the fiber size is 6 $\mu$m at the large end. A large LSO scintillator is attached to the front (larger) surface of the optical taper. Since the sampling area is small for the ICCD camera even with the optical taper compared to an x-ray film or image plate, we collected the preshot static diffraction pattern on a large film [Fig.~13(a)]. Comparing the static diffraction pattern on a larger detector and the dynamic pattern on a smaller CCD (subset of the former) is helpful for indexing diffraction peaks/spots.

%%%Fig.8
\begin{figure}[t]
\includegraphics[scale=0.6,clip]{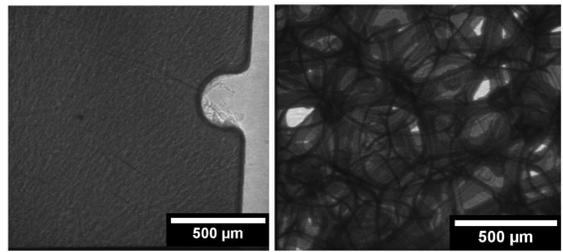}
\caption{Static PCI images of a trenched acetaminophen (Tylenol) specimen and low density Ni foam (0.45 g\,cm$^{-3}$; 20 pores per cm), obtained with a PI-Max ICCD camera. The sample thicknesses are about 9 mm in both cases. Undulator gap: 30 mm; scintillator: LuAG.  
}
\end{figure}

%%Fig9
\begin{figure}[t]
\includegraphics[scale=0.6,clip]{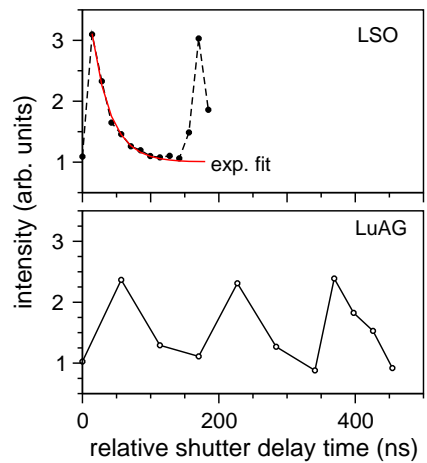}
\caption{Normalized intensity vs. relative shutter delay time for LSO and LuAG. The exponential fit ($y=\exp\{-x/\tau\}$ where $\tau$ detones the decay time) for LSO yields an effective decay time $\tau$=28 ns. }
\end{figure}

%%Fig10
\begin{figure}[ht]
\includegraphics[scale=0.5,clip]{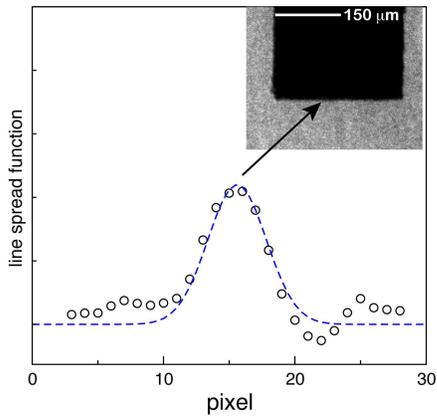}
\caption{Gaussian line spread function of the edge of a Cu rod (inset). The Gaussian function fit ($y=\exp\{-x^2/2\sigma^2\}$; the dashed line) yields $\sigma\approx2$ pixels, or 4 $\mu$m. }
\end{figure}

%%Fig11
\begin{figure}[t]
\includegraphics[scale=0.75,clip]{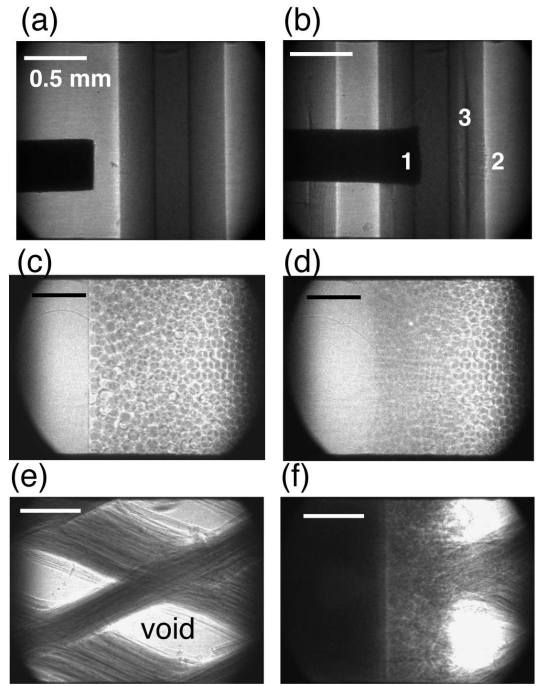}
\caption{Representative static (left column) and dynamic (right column) PCI images from APS shock experiments. The impact direction is from left to right. (a--b) Stainless steel plunger impact on vitreous carbon\cite{plunger} (shot \#11-49). (c--d) Borosillicate glass bead crushing\cite{glassbeads} (shot \#11-31). (e--f) Microlattice foam collapse (shot \#11-44). In (b), numbers 1--3 denote plunger deformation, and ejecta and spallation in vitreous carbon, respectively; cracks are induced around the plunger tip. Undulator gap: 26 mm or 30 mm; scintillator: LuAG.
}
\end{figure}

%%Fig12
\begin{figure}[t]
\includegraphics[scale=0.42,clip]{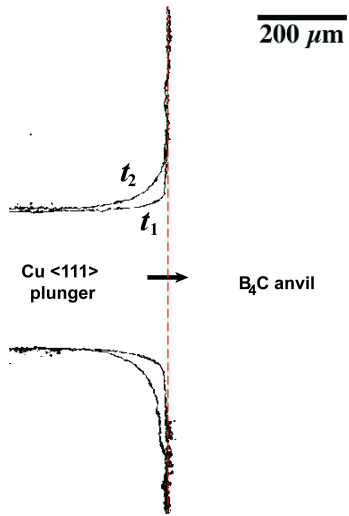}
\caption{Overlay of two dynamic PCI images of Cu $\langle111\rangle$ plunger impact on B$_4$C anvil at $t_1$ and $t_2$, recorded by ICCD1 and ICCD2 (the 2-frame ICCD camera, Fig.~7). $t_2-t_1$=459 ns. The processed images highlight the deformation profiles of the plunger. Shot \#12-24; undulator gap: 30 mm; scintillator: LuAG.
}
% and the projectile velocity$\sim$570 m\,s$^{-1}$. 

\end{figure}

%%%Fig13
\begin{figure}[ht]
\includegraphics[scale=0.45,clip]{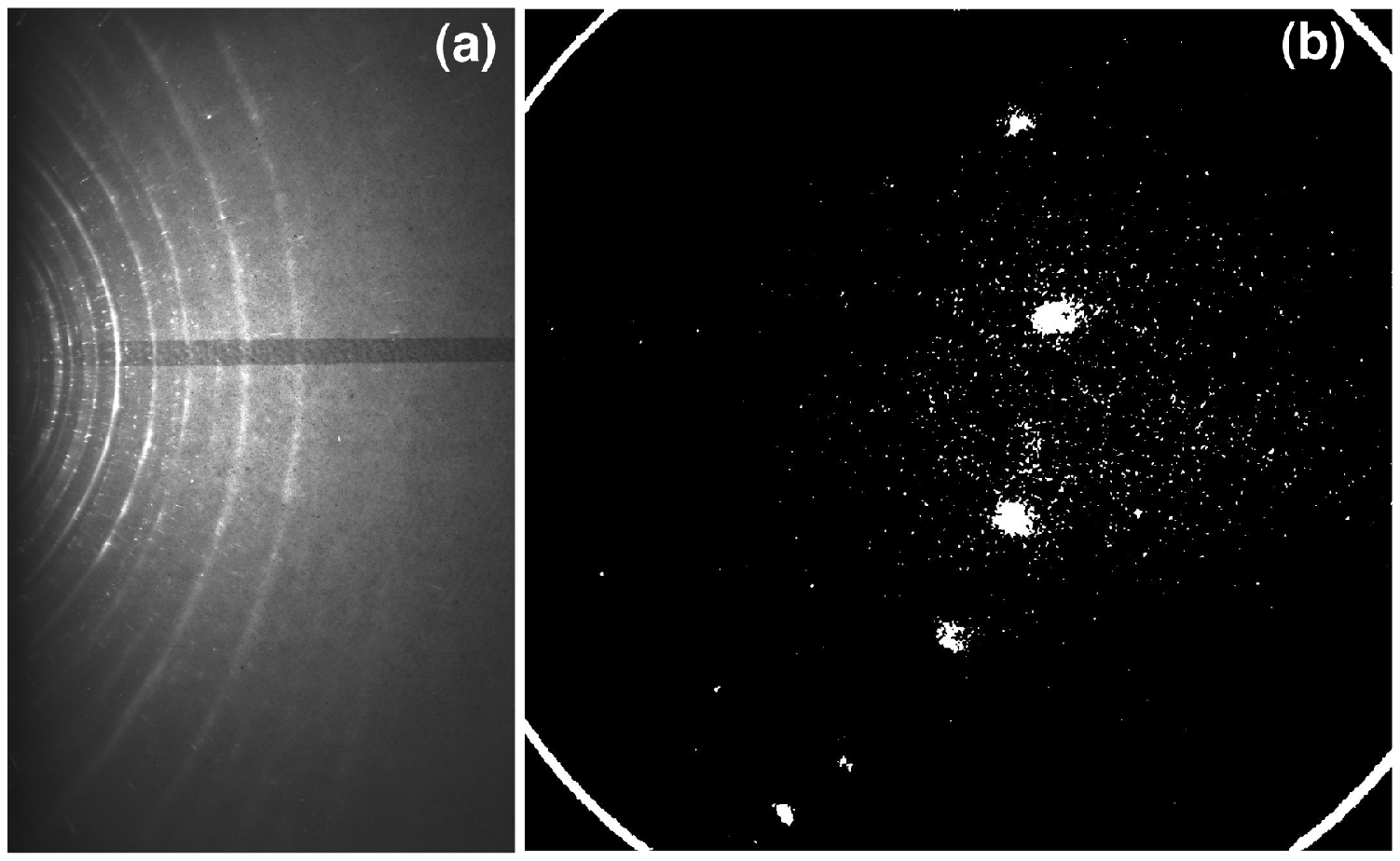}
\caption{(a) A static diffraction image of polycrystalline Fe recored by an x-ray film. (b) A dynamic Laue diffraction image of Fe (100) single crystal\cite{laue}
(shot \#11-52; undulator gap: 17 mm; scintillator: LSO).
}
\end{figure}

%%%\subsection{Scintillators and cameras}
Since scintillators are indispensable for our measurements with optical cameras, we compare two types of scintillators: 100 $\mu$m thick LuAG and 130 $\mu$m thick LSO/LYSO (Fig.~9). LYSO is similar to LSO in performance. To measure the decay curves of LSO and LuAG, we fixed the camera gate width at 20 ns, increased the camera delay incrementally relative to the RF pulse trigger and took PCI images. The image intensity vs. the delay time curves are plotted in Fig.~9. An exponential fit to the LSO data yields an effective decay time $\tau$=28 ns, consistent with, but slightly smaller than, the literature values.\cite{nikl06mst} However, the LuAG data points are too sparse for fitting. Qualitatively, Fig.~9 shows that the decay time of LuAG is longer, consistent with previous observations (e.g., 60 ns for LuAG vs. 30 ns for LSO).\cite{nikl06mst} Thus, LSO is preferred for its much shorter decay time. Nonetheless, the single-pulse PCI signal intensity obtained with LSO is about 4$\times$ lower than that with LuAG, so LuAG was used in most of our measurements. The longer decay time of LuAG may lead to the memory effect, i.e., the signals due to two x-ray pulses may overlay, resulting in a ``ghost'' image [Fig.~11(b)]. Caution should be exercised in data interpretation regarding this effect, although the ghost images could be useful in some cases.        

Most of our single-pulse PCI measurements were conducted with PI-MAX ICCD cameras and a 10$\times$ objective lens (Fig.~7). The spatial resolution as measured with an Xradia resolution pattern is $\sim$2 $\mu$m. The spatial resolution for a real specimen can be measured from a Gaussian line spread function of an edge. For example, we obtain the edge response function of a Cu rod sample with a machined edge from its PCI image, and the spatial derivative is the line spread function (Fig.~10). A Gaussian fit to the line spread function yields a root-mean-squared or RMS resolution of 4 $\mu$m. The ultimate spatial resolution of PCI at 32ID-B is expected to be 1 $\mu$m.\cite{shen07nima} 

The temporal resolution of PCI or diffraction is the x-ray pulse width, i.e., $\sim$80 ps.\cite{miceli_pilatus} In multiframe measurements, another relevant parameter is the minimum frame separation defined by the x-ray pulse separation (153 ns in the APS standard mode), although the framing rate of a framing camera can be faster or slower than the x-ray pulse repetition rate.

\subsection{Dynamic PCI and diffraction measurements}

Gas gun shock experiments with dynamic, single x-ray pulse, PCI and diffraction measurements  were conducted on representative materials and processes with single- or two-frame PI-MAX ICCD cameras, for different flyer plates and projectile velocities ($\sim$300--700 m\,s$^{-1}$). The camera gate width was set to be the pulse separation (153 ns). LuAG scintillators were used in PCI and LSO in diffraction experiments. We present some data to illustrate the capabilities of our experimental platform (Figs.~11--13); further details and scientific implications will be presented elsewhere. \cite{jensence,plunger,glassbeads,laue}

\begin{table*}[ht]
\begin{centering}
\caption{List of selected experiments and potential applications.}
\begin{tabular}{ll}
\hline\hline
\multicolumn{1}{c}{Dynamic experiments} & \multicolumn{1}{c}{Potential applications}\\\hline
PCI, Ce jet         & Jets, ejecta, flow and instability growth in solids, liquids, gases, and plasma.\cite{asay76apl,zhang98prl,dimonte11prl}\\
PCI, plunger & Ballistics; dynamic indentation; penetration; cracking; failure; spall; dynamic friction.\cite{meyers94b,kanel10ijf}\\
PCI, microlattice & Void nucleation or collapse; compression of low density materials, including \\
                  & plastics, metallic foams and gels; equation of state.\cite{lopatnikov04ijie,kanel10ijf,zhu07prb,arman10prb}\\
PCI, glass beads  & Compression of porous and granular materials; hotspots; powder reactions.\cite{meyers94b,thadhani93pms,huang12jap,bowden52b,an11prb}\\
Laue diffraction, Fe & Phase changes;\cite{klantar05prl} plasticity;\cite{ice11sci,turneaure09jap} concurrent diffraction and PCI measurements. \\  
\hline\hline
\end{tabular}
\end{centering}
\end{table*}

The PCI measurements of the impact of a stainless steel plunger (backed by a Lexan sabot) on a vitreous carbon anvil reveal rich impact phenomena [Figs.~11(a) and 11(b)], including plastic deformation of the plunger tip, spallation inside the anvil and ejecta on the anvil free surface. The compaction of granular materials was examined with PCI of 106 $\mu$m diameter borosillicate glass beads embedded in two polymethyl methacrylate (PMMA) plates [Figs.~11(c) and 11(d)]. Impact-induced void collapse in a microlattice plastic foam was also demonstrated with PCI [Figs.~11(e) and 11(f)]. 

We also obtained the static and dynamic PCI images (not shown) and observed the formation of Ce jets from the pre-engineered grooves upon plate impact. Some dynamic images show the memory effect of the LuAG scintillator. This memory effect may or may not be present, depending on the relative position between the camera gate width and the x-ray pulses or the corresponding scintillator decay curve. If the camera gate coincides with the maximum of the scintillator decay curve (there is a decay curve for each x-ray pulse), the memory effect due to the preceding pulse is minimum in the image of the current x-ray pulse. Otherwise, the memory effect is more pronounced. Such scenarios occur randomly unless the camera gate is synchronized with the x-ray pulses. 

The Taylor-impact PCI measurements (plunger impact, Fig.~12) with the 2-frame ICCD camera demonstrate the multiframe capability using multiplexed ICCD cameras. Besides PCI, single-pulse dynamic Laue diffraction measurement on an Fe (100) crystal was obtained during its impact on a vitreous carbon x-ray window, in the backscattering diffraction geometry [Fig.~13(b)].  

%%%%%%%%%%%%%%%%%%%%%%
%%%%%%%%%%%%%%%%%%%%%%

\section{Scientific opportunities and technical challenges}

Synchrotron x-ray sources provide high photon energy, high photon fluxes, high pulse repetition rates and sufficient coherency. Our synchrotron gas gun experiments with single-pulse x-ray PCI and diffraction measurements on representative materials and processes (Figs.~11--13) have demonstrated the potential of synchrotron-based shock platforms to address fundamental shock physics problems, and also revealed some technical issues. We discuss below some scientific opportunities and challenges. 

Due to its edge-enhancement capability, x-ray PCI is particularly advantageous for revealing structural inhomogeneities, especially in x-ray transparent or translucent materials (e.g., low-$Z$ materials), which are normally difficult to resolve with standard x-ray or proton radiography.\cite{yeager12comp} PCI is also useful for resolving the interfaces between an opaque and a translucent material. The current spatial resolution of x-ray PCI (2--4 $\mu$m; it can be improved to 1 $\mu$m) and temporal resolution of $<$100 ps are unprecedented for bulk-scale shock experiments. Thus, x-ray PCI can serve as an important complement to the well established yet still improving, highly penetrating, multiframe proton radiography,\cite{morris12ropp,king99nima,dimonte11prl} in particular for investigating mesoscale materials dynamics. The preliminary ``white beam'' x-ray diffraction measurement also shows its promise in revealing the atomic-level structures during dynamic loading. Table~II summarizes the potential applications of some dynamic experiments to shock physics.

However, technical issues or challenges do exist in detection, light source and dynamic loading (they are coupled) and require appreciable development efforts, including

\begin{itemize}
\item {\it Scintillators.} For dynamic PCI measurements, it appears that the immediate  issue to be resolved is scintillator darkening and damage related to heat load imposed by the x-rays. The APS photon fluxes are not exhausted yet, since the undulator gap is mostly 26--30 mm in our experiments (it can be reduced to 20 mm, and the flux will increase by a factor of 3 from the 30 mm gap; Fig.~2). The decay time of LSO or LYSO is satisfactory, but its light yield and damage resistance are not quite. One way to improve the yield is microstructure engineering (e.g., using columnar structure).\cite{nikl06mst} LuAG is still useful, in particular if we reduce the scintillator exposure time (or the x-ray time window and thus heat load) and synchronize the x-ray pulse with the camera gate (to reduce the memory effect). Exploring efficient, high damage resistance, fast decay scintillators is the first priority. K$_2$LaI$_5$:Ce is a possible candidate.\cite{nikl06mst} For diffraction, the scintillator damage is of less concern.  
After the scintillator issue is resolved, we can apply the phase retrieval process\cite{paganin02jm} to the PCI images, which may yield more quantitative information such as the mass density. 
%%and YSO (Y$_2$SiO$_5$).

\item {\it Timing and synchronization.} Reducing the jitter in the projectile launch can reduce the x-ray time window and improve synchronization of the x-ray time window  with dynamic events. For example, instead of using the gas gun firing signal, a photodiode or electromagnetic sensor could be placed in the projectile path to supply the trigger to the slow shutter. A fast, heat resistant shutter system can be designed to open-close the x-ray beam when triggered by the impact pin signal. Using a different loading device for bulk-scale shock experiments, e.g., a powder gun, can also reduce the jitter.

\item {\it Multiframe 2D imaging/diffraction detectors.} In the best case, the number of photons currently available may allow 1-to-4 beam splitting for multiframe recording with optical ICCD cameras, and the optical intensifier response spectrum should be optimized for the scintillator used. X-ray intensifiers should also be considered. To eliminate the scintillator and beam splitting, direct-detection, high quantum efficiency, hybrid CMOS cameras are likely the solution. Improving, designing and fabricating such cameras with high sensitivity and high framing rate are necessary, although there are concerns regarding x-ray beam manipulation for PCI and high flux x-ray damage to the cameras. 

\item {\it Diffraction.} Large diffraction detection area is highly desirable for structure refinement, and high fluxes in harder x-rays are useful for deeper penetration depth. 

\item {\it Probing x-ray beam: higher fluxes, larger beam sizes, multibeams, and controllability.} Shock experiments are or will become x-ray photon-starved so higher flux capability is always beneficial. Multibeam PCI would allow at least two view angles or enable limited tomography. Larger beam size or field of view can be beneficial for PCI and direct-detection cameras. Improved controllability includes the flexibility in the bunch time structure.

\item{\it Targeted science.} Targeted experiments should be designed within the limitations of the x-ray source and detectors. For increased penetration depth, proton radiography can be used to complement x-ray PCI.

\end{itemize}

\section{Summary}
We have established the feasibility of gas gun shock experiments with dynamic x-ray PCI and diffraction measurements at APS. Ultrafast, single-pulse PCI measurements with unprecedented temporal ($<$100 ps) and spatial ($\sim$2 $\mu$m) resolutions for bulk-scale shock experiments have been demonstrated, as well as single-pulse Laue diffraction. Our results not only substantiate the potential of synchrotron-based shock platforms to address a variety of shock physics problems, but also highlight the technical challenges related to detection, x-ray source and dynamic loading.

\begin{acknowledgments}
We appreciate professional help from J. Esparza, C. Owens, T. Pierce (LANL), A. Deriy, and B. Glagola (ANL) in various stages of our experiments. We have benefited from valuable discussions with C. Barnes, W. Buttler, F. Cherne, G. Dimonte, D. Fulton, D. Funk, S. Greenfield, R. Martineau, D. Montgomery, R. Olson, D. Oro, R. Saavedra, J. Sarrao, D. Stahl, R. Valdiviez, Z. Wang (LANL), R. Hemley, G. Shen, Y. Xiao (HPCAT), J. Wang (APS), O. Tschauner (UNLV), and S. Gruner (Cornell). D. Dattelbaum is thanked for suppling the microlattice sample, and D. A. Fredenburg, for his contribution in the glass beads experiments. This work was partly supported by LANL's LDRD (LDRD-20110585ER), MaRIE and Science Campaign programs. LANL is operated by Los Alamos National Security, LLC for the U.S. Department of Energy under Contract No. DE-AC52-06NA25396. Use of the Advanced Photon Source, an Office of Science User Facility operated
for the US Department of Energy (DOE) Office of Science by
Argonne National Laboratory, was supported by the US DOE under
Contract No. DE-AC02-06CH11357.
\end{acknowledgments}

%\newpage
\bibliographystyle{apsrev4-1}
\bibliography{rsi-2ca}

\end{document}